\documentclass[10pt]{iopart}
\usepackage{epsfig,graphicx}

\usepackage[pdflatex=true,recompilepics=auto]{gastex}

\usepackage{color}

\newcommand{\bT}{\ensuremath{\mathbf{T}}}
\newcommand{\bP}{\ensuremath{\mathbf{P}}}
\newcommand{\bD}{\ensuremath{\mathbf{D}}}
\newcommand{\bM}{\ensuremath{\mathbf{M}}}
\newcommand{\bS}{\ensuremath{\mathbf{S}}}

\newcommand{\sgn}{\textrm{sgn}}

\newcommand{\Prob}{\mathrm{P}}
\newcommand{\W}{\mathrm{W}}
\newcommand{\Z}{\mathrm{Z}}

\begin{document}

\title[One-dimensional majority cellular automata with thermal noise]{Transfer matrix analysis of one-dimensional majority cellular automata with thermal noise}

\author{R\'emi Lemoy$^{1,2}$,  Alexander Mozeika$^2$ and Shinnosuke Seki$^2$}
\address{$^1$Department of Applied Physics,\\
$^2$Department of Information and Computer Science,\\
 Aalto University, FI-00076, Aalto, Finland.}




\ead{remilemoy@gmail.com, \{shinnosuke.seki, alexander.mozeika\}@aalto.fi}



\begin{abstract}
Thermal noise in a cellular automaton refers to a random perturbation to its function which eventually leads this automaton to an equilibrium state controlled by a temperature parameter.  We study the 1-dimensional majority-3  cellular automaton under this model of noise. Without noise, each cell in this automaton  decides its next state by majority voting among itself and its left and right neighbour cells. Transfer matrix analysis shows that the automaton always reaches a state in which every cell is in one of its two states with probability 1/2 and thus cannot remember even one bit of information. Numerical experiments, however, support the possibility of reliable computation for a long but finite time.
\end{abstract}



\section{Introduction\label{section:intro}}
In theoretical computer science, reliable computation in the presence of noise has been attracting researchers since von Neumann's work on the fault-tolerance of Boolean circuits in 1956 \cite{Neumann1956}; a faultless circuit was simulated by a circuit of {\em noisy} gates that is larger only by logarithmic factor and performs the same computation with some constant probability of success close to one.  In his simulation, the fault-tolerance mainly stems from {\em majority voting} and {\em redundancy}: each intermediate result is computed several times and its value is determined to be the one supported at least by half. Non-uniform connectivity among gates is also important in order to obtain fault-tolerance 
\cite{DobrushinOrtyukov1977,Pippenger1985,Spielman1996}. 

The possibility to manufacture systems with homogeneous components and advances in parallel computation call for the study of fault-tolerance in models with uniform structure. 
The cellular automaton (CA), introduced by Ulam \cite{Ulam1952} and von Neumann \cite{Neumann1966}, is one of them. A CA is a regular array of cells, where each cell can be in two or more states, and computes a function of its neighbouring cells at discrete time steps $t$. Fault-tolerance of CA has been investigated under various noise models \cite{Gacs1986,Gacs2001,McCannPippenger2008}, but mainly subject to the {\em flip noise} which, regardless of the computation,  affects the output of a cell (e.g. inverts its binary output with some probability $\epsilon$).

In statistical physics (SP), studies of CA go back to  Wolfram~\cite{ Wolfram1983}, who studied the behaviour of CA by simulations and analytically. The CA model of Wolfram  is a deterministic limit of more general probabilistic CA (PCA)~\cite{Gacs2001}. A large class of PCA satisfies the detailed balance condition for some Hamiltonian (energy) function~\cite{Grinstein1985}. Hence as $t\rightarrow\infty$, they are governed  by a stationary distribution of the  Gibbs-Boltzmann type. The question of fault-tolerance is then directly related to the occurrence  of \emph{phase transitions}, i.e. the presence of several stationary distributions~\cite{Pra2002}, in infinite size systems. Furthermore,  a more general  scenario can be treated by mapping non-equilibrium distributions of PCA into equilibrium SP models~\cite{Domany1984, Lebowitz1990}.

In this paper, we study the fault-tolerance of the 1-dimensional majority-3 cellular automaton (1d-MAJ3-CA) with \emph{thermal noise}. The 1d-MAJ3-CA is a 1-dimensional array of identical 2-state cells. The state of a cell is updated by taking the majority vote among itself and its left and right neighbour cells.  Thermal noise disturbs the computation in the gates by adding a random term proportional to the {\em temperature} parameter $T$.  Although the flip noise makes the result of a computation surely incorrect once it takes place, the thermal noise does nothing beyond competing with it, and hence, even if it occurs, the computational result can still be correct. Here, using the transfer matrix method~\cite{Baxter1982}, we show that, as long as the temperature is positive, the thermal noise eventually leads the 1d-MAJ3-CA to the equilibrium state in which each cell is in one of its two states with probability 1/2. In other words, it cannot ``remember one bit of information'' forever (Gacs discussed this 
problem for the flip noise~\cite{Gacs1986} and stated analogous results). 

This paper is organised as follows.  We define the model in section \ref{section:model}. In section \ref{section:MC}, we present numerical simulations, and use the transfer matrix method in section \ref{section:equilibrium} to study the equilibrium properties of this model. Finally, in section \ref{section:summary} we summarise and discuss our results.

\section{Model\label{section:model}}

The 1d-MAJ3-CA consists of $N$ cells interacting on the 1-dimensional lattice, with periodic boundary. 
The state of the $i$-th cell at time $t$ is denoted by the Ising variable $s_i(t) \in \{-1,1\}$. 
The {\em configuration} $s(t)=(s_1(t),\ldots,s_N(t))$ of the CA is the assignment of a state (-1 or 1) to each cell. 
The dynamics is governed by the synchronous stochastic alignment process
\begin{eqnarray}\label{def:process}
s_i(t+1)=\sgn\left[h_i(s(t))+T\eta_{i}(t)\right], 
\end{eqnarray}
where the {\em local field} $h_i(s(t))$ is given by $h_i(s(t))=s_{i-1}(t)+s_{i}(t)+s_{i+1}(t)$, and $\sgn$ is the sign function defined as $\sgn[x] = +1$ for $x \ge 0$ and $\sgn[x] = -1$ for $x < 0$. 
The noise, represented by $\eta_{i}(t)$, is competing with the local field. Its amplitude is controlled by the temperature parameter  $T\in [0,\infty)$. 
For $T=0$, the process (\ref{def:process}) converts into a deterministic majority-3 voting, while for $T\rightarrow \infty$ it is completely driven by  noise.

Let us assume that  $\eta_{i}(t)$ are independent and identically distributed random variables  coming from the distribution 
\begin{eqnarray} 
\Prob(\eta)=\frac{1}{2}\left[1-\tanh^2(\eta)\right].\label{def:noise}
\end{eqnarray}
Under this distribution of noise, the process (\ref{def:process}) leads us to the probability law
\begin{eqnarray}
\Prob[ s_i(t+1)\vert s(t)]=\frac{1}{2}\left[1+ s_i(t+1)\tanh\left(\beta h_i(s(t))\right)\right],\label{eq:average-process}
\end{eqnarray}
where $\beta=1/T$ is the inverse temperature. This model of noise was used in studies of neural networks~\cite{Peretto1984, TheorOfIP}, and of metastablity in PCA ~\cite{Bigelis1999}.

Given a configuration $s(t)$ at time $t$, the configuration $s(t+1)$ at time $t+1$ is governed by the probability $\W[s(t+1)\vert s(t)]=\prod_{i=1}^N \Prob[ s_i(t+1)\vert s(t)]$. This is a transition probability of the Markov process 
\begin{eqnarray}
\Prob_{t+1}(s) &=& \sum_{\hat s} \W[s\vert \hat s] \Prob_{t}(\hat s),\label{eq:master}
\end{eqnarray}
which traces the evolution of the probability $\Prob_{t}(s)$ to find the variables in a state $s$ at time $t$.

The choice (\ref{def:noise}) for the noise distribution guarantees that the system  (\ref{eq:master}) has a unique~\cite{Peretto1984} stationary solution  
\begin{equation}
\Prob_{\infty}(s)=\frac{1}{\Z}\prod_{i=1}^{N} 2\cosh(\beta h_i(s)) ,\label{eq:equilibrium1}
\end{equation}
where $\Z=\sum_s \prod_{i=1}^{N} 2\cosh(\beta h_i(s)) $ is the partition function.  This distribution  can be also written in a more familiar (equilibrium) Gibbs-Boltzmann form $\Prob_{\infty}(s)=\mathrm{e}^{-\beta E(s)}/\Z$ upon definition of the ``energy'' function $E(s)=-\frac{1}{\beta}\sum_{i=1}^{N}\log2\cosh(\beta h_i(s))$ (this is not a proper energy function, or Hamiltonian, because of its explicit dependence on the noise parameter $\beta$). Furthermore, for the probabilistic majority-3 automaton, the choice of (\ref{def:noise}) is the only type of noise which guarantees the convergence to Gibbs-Boltzmann equilibrium \cite{Grinstein1985}. 

Finally, let us compare the thermal noise model (\ref{eq:average-process}) with the flip-noise model 
\begin{eqnarray}
\Prob[ s_i(t+1)\vert s(t)]=\frac{1}{2}\left[1+ s_i(t+1)\,\sgn\left[h_i(s(t))\right] \tanh(\tilde \beta)                                 \right],\label{eq:average-process-for-flip-noise}
\end{eqnarray}
which corresponds to  the process $s_i(t+1)=\xi_{i}(t)\,\sgn\left[h_i(s(t))\right]$, where  $\xi_i(t) \in \{-1,1\}$ and $\Prob(\xi_{i}(t)=-1)=[1-\tanh(\tilde \beta)]/2$. The probability of an error ($s_i(t+1)\,\sgn\left[h_i(s(t))\right] =-1$) in this model, and in the thermal noise model, is given by the functions $[1-\tanh(\tilde \beta)]/2$ and $[1-\tanh(\beta\vert h_i(s(t)) \vert)]/2$, respectively. Suppose that one of these models is in the ordered state, and the probability of an error in the other model is not higher than in this one. Then intuitively, both models are in the ordered state. This intuition turns out to be correct for models with non-uniform topologies~\cite{Mozeika2010, Mozeika2011}, but whether this is true in a more general case is not clear.

\section{Monte-Carlo simulations \label{section:MC}} 
	
\begin{figure}[t]
\begin{minipage}{0.45\linewidth}
\begin{center}
\includegraphics[scale=0.45]{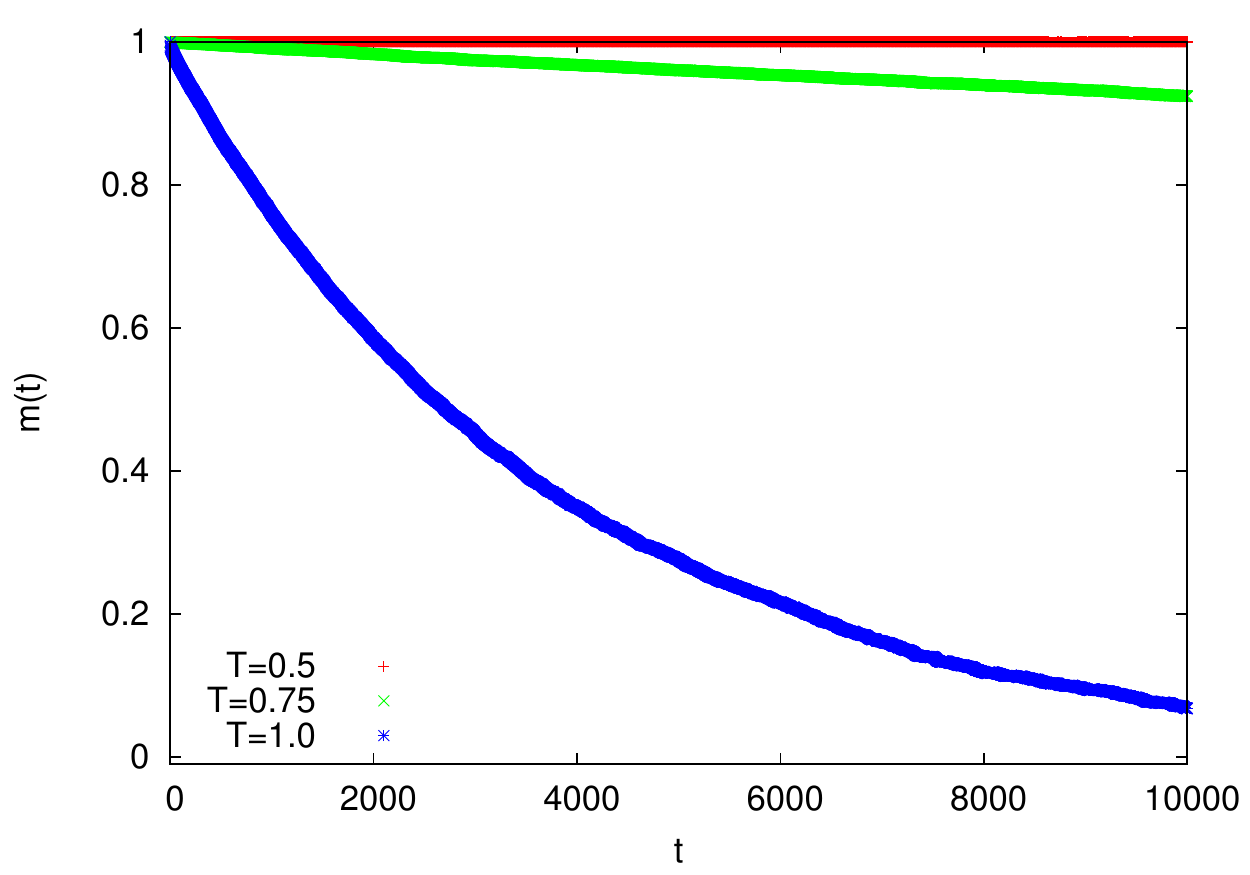}
\end{center}
\end{minipage}
\begin{minipage}{0.45\linewidth}
\begin{center}
\includegraphics[scale=0.45]{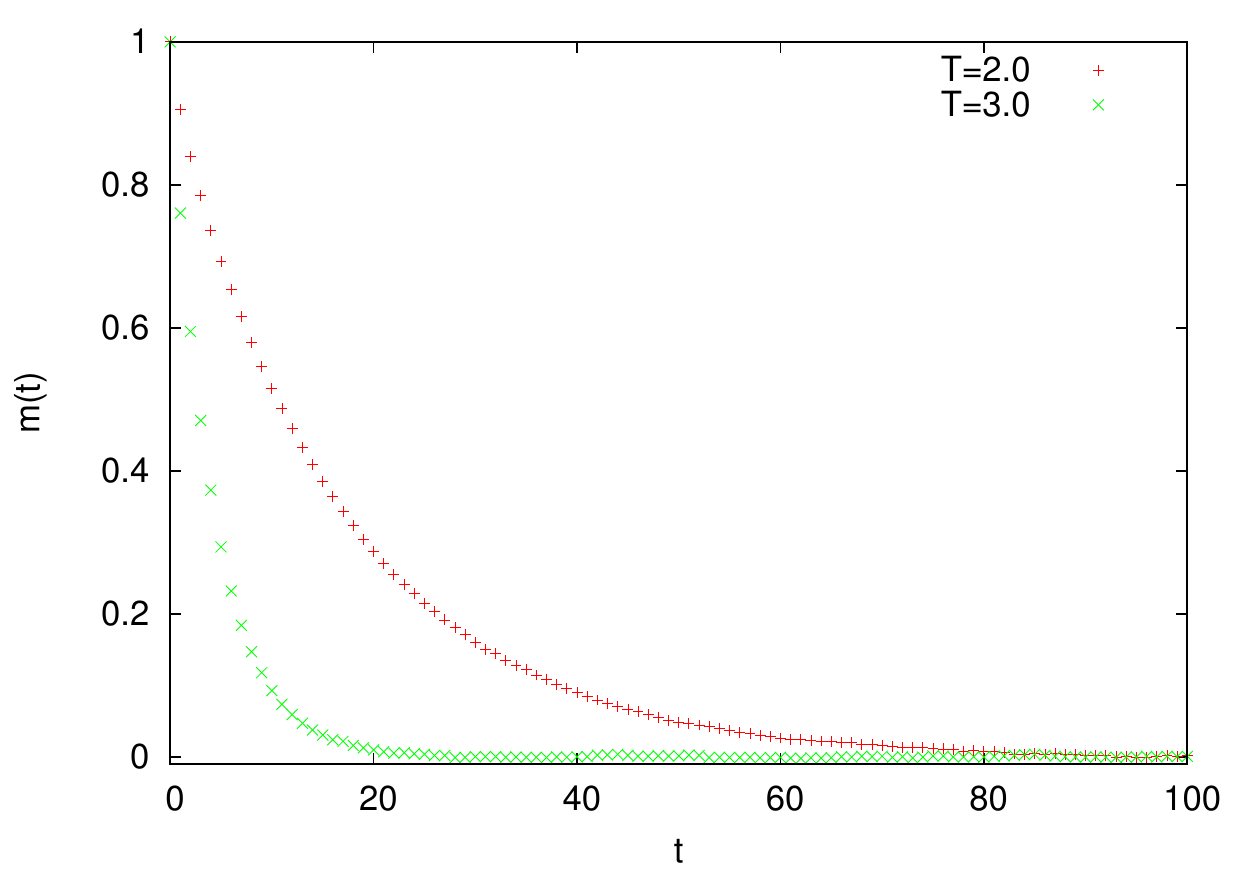}
\end{center}
\end{minipage}
\caption{Remembering 1-bit of information by the 1d-MAJ3-CA automaton  with $N=10^6$ cells in the presence of noise with temperature $T$. Left: $T = \{0.5, 0.75, 1.0\}$  (from top to bottom).  Right: $T = \{2.0, 3.0\}$  (from top to bottom). }
\label{fig:MC}
\end{figure}
In this section, we show with Monte-Carlo (MC) simulations of the process (\ref{def:process}), starting from the configuration with all cells in the state $+1$, that the 1d-MAJ3-CA automaton of finite size can remember this one bit of information for a considerable amount of time, even at relatively high temperatures. As McCann and Pippenger point out in~\cite{McCannPippenger2008}: ``this property of remembering a bit is all that is needed to achieve fault-tolerant computation.''

In Figure~\ref{fig:MC}, we plot the \emph{magnetization} $m(s(t)) = \frac{1}{N}\sum_{i=1}^{N} s_i(t)$ measured in the MC simulation of $N = 10^6$ cells evolving in time $t$ for low and high temperatures. Here the state  $m(s(t)) = 1$ means that the system holds 1-bit of information by having all its cells in the state $+1$, whereas if $m(s(t))=0$, then half of the cells are in $+1$ and the other half are in $-1$, i.e., the information is lost. We note that even after $10^4$ time-steps, more than $90\%$ of the cells of the 1d-MAJ3-CA are in the correct state $+1$ at $T = 0.75$.   Further analysis  of the  dynamics is required  in order to say how quickly  the  information about the initial state of the automaton gets lost.

\section{Transfer matrix analysis of the equilibrium \label{section:equilibrium}}

The equilibrium probability distribution  (\ref{eq:equilibrium1}) can be used to compute thermal averages such as $\langle s_i\rangle= \sum_s\Prob_{\infty}(s)s_i$ and $\langle s_is_j\rangle= \sum_s\Prob_{\infty}(s)s_is_j$. The former will allow us to show that the 1d-MAJ3-CA indeed forgets its initial configuration after a long time, and the latter will allow us to study correlations. Let us first consider the partition function
\begin{eqnarray}
\label{eq:partition}
Z &= & \sum_{s_1, \ldots, s_N } \prod_{i=1}^{N} 2\cosh(\beta(s_{i-1} + s_{i} + s_{i+1}))\label{def:Z}
\end{eqnarray}
where $s_0=s_N$ and $s_1=s_{N+1}$ (periodic or closed boundary).  In order to compute this  function, we will use the transfer-matrix  method~\cite{Baxter1982}.

\begin{figure}
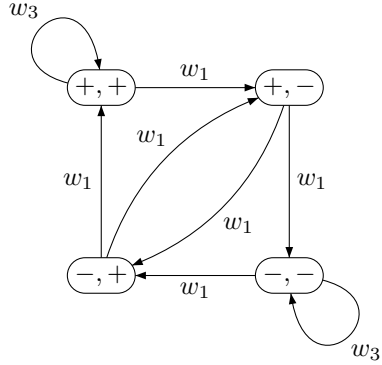

\begin{center}
\begin{gpicture}(40,40)(0,0)

\node[Nadjust=wh](A11)(5,30){$+, +$}
\node[Nadjust=wh](A10)(30,30){$+, -$}
\node[Nadjust=wh](A00)(30,5){$-, -$}
\node[Nadjust=wh](A01)(5,5){$-, +$}

\drawloop[loopangle=135](A11){$w_3$}\drawedge(A11,A10){$w_1$}
\drawedge(A10,A00){$w_1$}\drawedge[curvedepth=5](A10,A01){$w_1$}
\drawloop[loopangle=315](A00){$w_3$}\drawedge(A00,A01){$w_1$}
\drawedge(A01,A11){$w_1$}\drawedge[curvedepth=5](A01,A10){$w_1$}

\end{gpicture}
\end{center}

\caption{Transition diagram of the auxiliary two-cell binary-state automaton, with $w_3 = 2\cosh(3\beta)$ and $w_1 = 2\cosh(\beta)$.}
\label{fig:maj3_graph}
\end{figure}

To this end, we consider an auxiliary two-cell binary-state automaton, whose state diagram  is presented in Figure \ref{fig:maj3_graph}.  This diagram gives rise to the "transition" matrix
\begin{eqnarray}
\bT=\left( \begin{array}{cccc}
2\cosh(3\beta) & 2\cosh(\beta) & 0 & 0 \\
0 & 0 & 2\cosh(\beta) & 2\cosh(\beta) \\
2\cosh(\beta) & 2\cosh(\beta) & 0 & 0 \\
0 & 0 & 2\cosh(\beta) & 2\cosh(3\beta)  \end{array} \right),
\end{eqnarray}
where rows and columns are labelled by  the binary vectors $(+1;+1)$, $(+1;-1)$, $(-1;+1)$ and $(-1;-1)$, in this order.  An element of this matrix  is defined by $\bT[s_i,s_j\vert s_k,s_l]=2\cosh(\beta(s_i+s_j+s_l))\delta_{s_j; s_k}$, where $\delta_{s_j; s_k} $ is the Kronecker delta.  After $N$ ``time-steps'', the state of  this automaton is governed by the operator $\bT^N$. Now due to the properties of the matrix $\bT$,  we have
\begin{eqnarray}
\Tr(\bT^N)&=&\sum_{s_1,...,s_N}\sum_{s_1^{\prime},...,s^{\prime}_N}\prod_{i=1}^{N}\bT[s_{i-1},s_i^{\prime}\vert s_i, s_{i+1}^{\prime}]
\\
&=&\sum_{s_1,...,s_N} \prod_{i=1}^{N}\bT[s_{i-1},s_i\vert s_i, s_{i+1}]=Z\nonumber.
\end{eqnarray}
Thus $\bT$ is a transfer matrix for the original majority-3 system. This construction can be used to obtain a transfer matrix for \emph{any} one-dimensional majority voting system where the vote concerns $2k+1$ cells (the central cell and $k$ neighbours to the left and to the right of this cell). However, the dimensionality of the  transfer matrix grows exponentially with $k$ (see \ref{section:Maj2k}).

For majority-3, we note first that the transfer matrix $\bT$ has four distinct eigenvalues and can be diagonalised as $\bP^{-1}\bT\bP = \bD$. $\bD$ is a diagonal matrix constructed from the eigenvalues
\begin{eqnarray}
\lambda_{1,3} &=& \frac{1}{2} \left(w_3 + w_1 \pm \sqrt{(w_3-w_1)^2  + 4 w_1^2}\right)\label{eq:eigenv}\\
\lambda_{2,4} &=& \frac{1}{2} \left(w_3 - w_1 \pm \sqrt{(w_3+w_1)^2  - 4 w_1^2}\right) \nonumber
\end{eqnarray}
 of $\bT$,  where $w_3 = 2\cosh(3\beta)$ and $w_1 = 2\cosh(\beta)$. $\bP$ is the matrix whose columns are the corresponding eigenvectors of $\bT$. Then using the properties of the trace,  we obtain $Z= \lambda_1^N + \lambda_2^N + \lambda_3^N + \lambda_4^N$.
In the thermodynamic limit  $N \to \infty$, the partition function $Z$ is dominated by the largest eigenvalue $\lambda_1=\left(w_3 + w_1 +  \sqrt{(w_3-w_1)^2  + 4 w_1^2}\right)/2$ for any finite $\beta$. 

Let us now compute local magnetisations and correlations. To this end, we use the relation
\begin{equation}
\label{invariance}
P_\infty(s)=\sum_{\tilde{s}}W[s|\tilde{s}]P_\infty(\tilde s), 
\end{equation}
which expresses the invariance of the equilibrium distribution. All sites are equivalent because of the periodic boundary, hence it is sufficient to consider the average magnetization $\left<s_1\right>=\sum_{s}P_\infty(s)s_1$ at site $1$. Using the identity (\ref{invariance}), we obtain
\begin{equation}
\label{site1}
\left<s_1\right>=\sum_{\tilde{s}}\tanh(\beta h_1(\tilde{s}))P_\infty(\tilde{s}).
\end{equation}
Now
\begin{eqnarray}
\left<s_1\right>&=&\frac{1}{Z} \sum_s \tanh \left(\beta h_1(s)\right)\prod_{i=1}^N 2\cosh \left(\beta h_i(s)\right)\nonumber\\
&=&\frac{1}{Z} \sum_s 2\sinh\left(\beta h_1(s)\right)\prod_{i=2}^N 2\cosh\left( \beta h_i(s)\right) \nonumber \\
&=&\frac{\Tr(\bS\bT^{N-1})}{\Tr(\bT^N)},\label{eq:s1}
\end{eqnarray}
where
\begin{eqnarray}
\bS=\left( \begin{array}{cccc}
2\sinh (3\beta) & 2\sinh (\beta) & 0 & 0 \\
0 & 0 & 2\sinh (\beta & -2\sinh (\beta) \\
2\sinh (\beta) & -2\sinh (\beta) & 0 & 0 \\
0 & 0 & -2\sinh (\beta) & -2\sinh (3\beta)  \end{array} \right).
\end{eqnarray}
%
Finally, defining the matrix $\bM=\bP^{-1}\bS\bP$, we obtain
\begin{equation}
\label{finalmag}
\left<s_1\right>= \frac{\sum_{i=1}^4 \bM_{i,i}\lambda_i^{N-1}}{\sum_{i=1}^4 \lambda_i^N}.
\end{equation}
In the thermodynamic limit, $\lim_{N\to\infty}\left<s_1\right>=\bM_{1,1}/\lambda_1$. However, $\bM_{1,1}=0$ (see \ref{mat}) and thus $\left<s_i\right>=0$ for all $i$ at any finite $\beta$, i.e. the system is in a disordered (paramagnetic) state.

Let us now compute the correlations $\left<s_is_j\right> =\sum_{s}P_\infty(s)s_is_j$ for $j>i$. Using the equilibrium relation (\ref{invariance}) and following the steps of (\ref{eq:s1}), we obtain
\begin{eqnarray}
\left<s_is_j\right>&=& \frac{\Tr(\bT^{i-1}\bS\bT^{j-i-1}\bS\bT^{N-j})}{\Tr(\bT^N)}\nonumber\\
&=& \frac{\Tr(\bM\bD^{j-i-1}\bM\bD^{N+i-j-1})}{\Tr(\bT^N)}\nonumber\\
&=&\frac{\sum_{k,l=1}^4 \bM_{k,l}\lambda_l^{j-i-1}\bM_{l,k}\lambda_k^{N+i-j-1}}{\sum_{k=1}^4 \lambda_k^N}
\end{eqnarray}
In the thermodynamic limit $N\to\infty$, this equation reduces to 
\begin{eqnarray}
\left<s_i s_j\right>&=&\sum_{k=1}^4 \frac{\bM_{1,k}\bM_{k,1}}{\lambda_1^2}\left(\frac{\lambda_k}{\lambda_1}\right)^{j-i-1}\nonumber\\
&=&\left<s_i\right>\left<s_j\right>+\sum_{k=2}^4 \frac{\bM_{1,k}\bM_{k,1}}{\lambda_1^2}\left(\frac{\lambda_k}{\lambda_1}\right)^{j-i-1}.
\end{eqnarray}
However, $\bM_{1,3} =0=\bM_{3,1}$ (see \ref{mat}) and $\left<s_i\right> =0$ for all $i$. Thus 
\begin{eqnarray}
\left<s_i s_j\right>&=&     \frac{\bM_{1,2}\bM_{2,1}}{\lambda_1 \lambda_2}\left(\frac{\lambda_2}{\lambda_1}\right)^{j-i} + \frac{\bM_{1,4}\bM_{4,1}}{\lambda_1 \lambda_4}\left(\frac{\lambda_4}{\lambda_1}\right)^{j-i}.
\end{eqnarray}
Furthermore, it can be shown that 
\begin{eqnarray}
  \frac{\bM_{1,2}\bM_{2,1}}{\lambda_1 \lambda_2} + \frac{\bM_{1,4}\bM_{4,1}}{\lambda_1 \lambda_4}=1,
\end{eqnarray}
so we denote 
\begin{eqnarray}
\alpha =  \frac{\bM_{1,2}\bM_{2,1}}{\lambda_1 \lambda_2}
\end{eqnarray}
and write
\begin{eqnarray}
\left<s_i s_j\right>&=&    \alpha \left(\frac{\lambda_2}{\lambda_1}\right)^{d} + (1-\alpha)\left(\frac{\lambda_4}{\lambda_1}\right)^{d}, 
\end{eqnarray}
where $d=j-i$.

\begin{figure}[tb]
\begin{center}
\begin{minipage}{0.45\textwidth}
\includegraphics[scale=0.3]{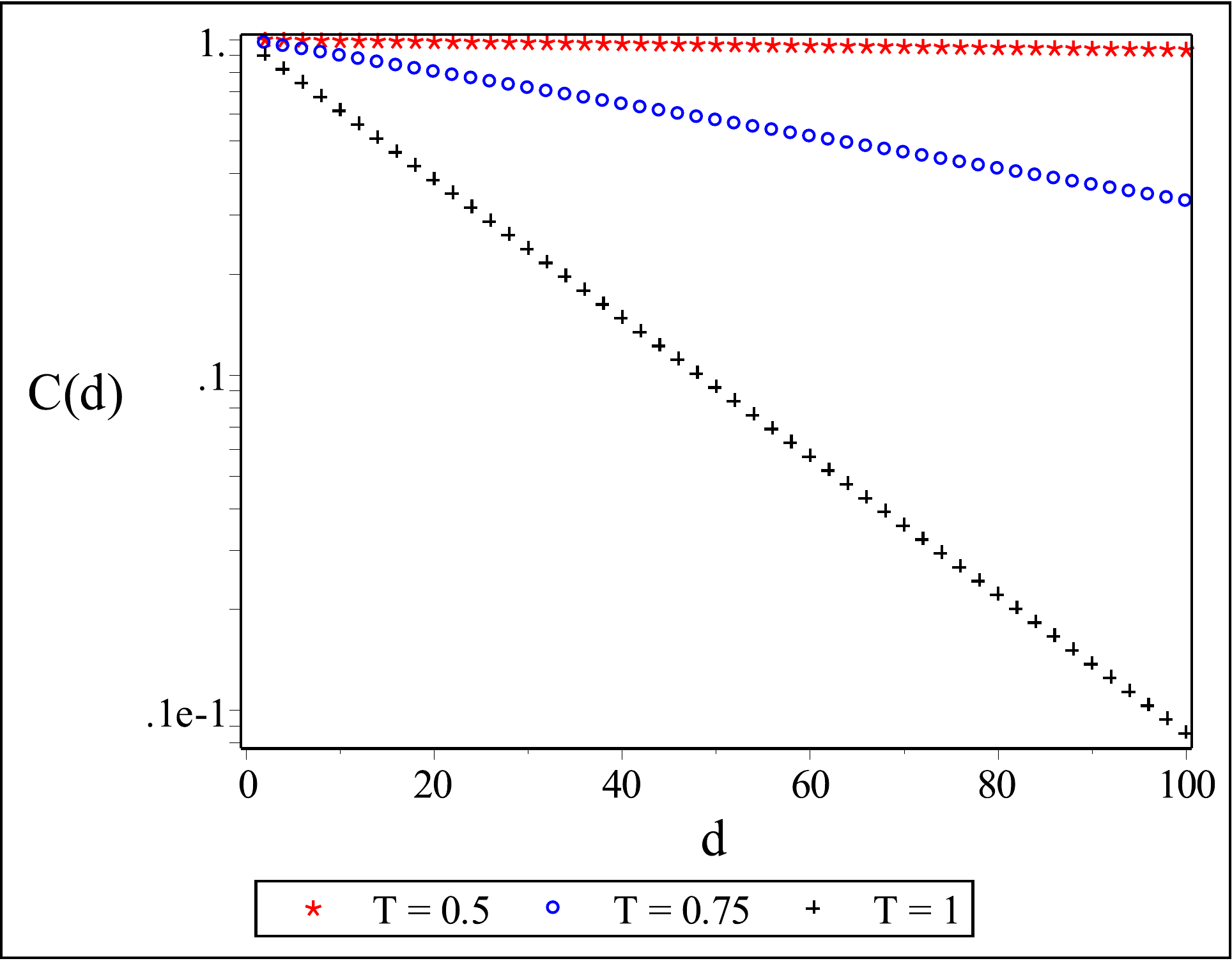}
\end{minipage}
\begin{minipage}{0.45\textwidth}
\includegraphics[scale=0.3]{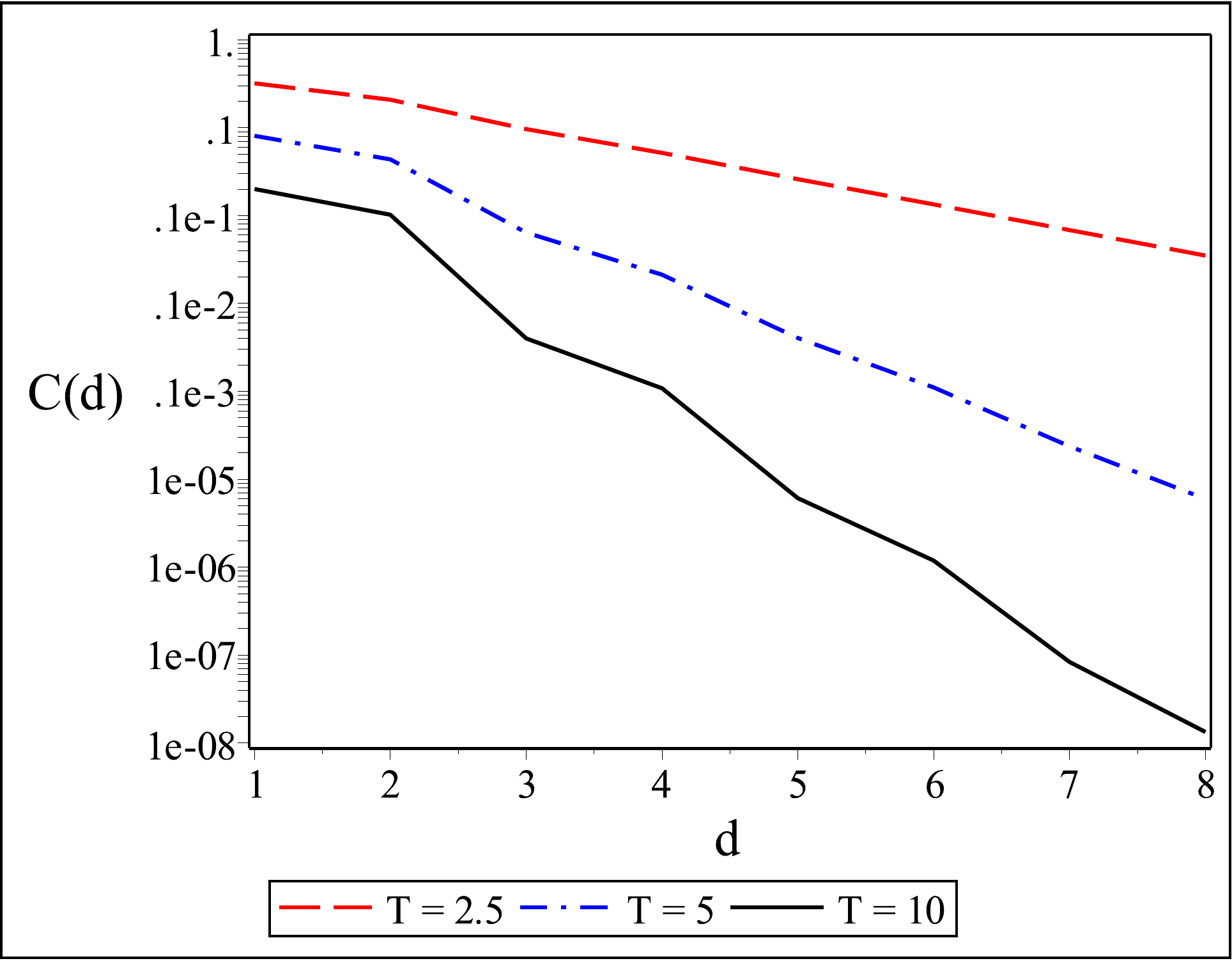}
\end{minipage}
\caption{Behaviour of the correlation function $C(d)= \langle s_0s_d\rangle$ in the 1d-MAJ3-CA predicted by the theory for high and low temperatures $T$.  Left: $T = \{0.5, 0.75, 1\}$.  Right: $T = \{2.5, 5, 10\}$. }
\label{fig:correlation0}
\end{center}
\end{figure}

In the high temperature limit $\beta \to 0$, the ratios $\lambda_2/\lambda_1$, $\lambda_4/\lambda_1$ and the function $\alpha$ behave as follows: 
\begin{eqnarray}
\frac{\lambda_2}{\lambda_1} &=& \beta +\Or(\beta^2),\nonumber\\
\frac{\lambda_4}{\lambda_1} &=& -\beta +\Or(\beta^2),\nonumber\\
\alpha&=&     \frac{1}{2}+\beta+\Or(\beta^3).
\end{eqnarray}
Thus in this limit,
\begin{eqnarray}
\left<s_i s_j\right>&=&\gamma_d(\beta +\Or(\beta^2))^d     
\end{eqnarray}
where $\gamma_d=1+\Or(\beta^3)$ if $d$ is even and $\gamma_d=\beta+\Or(\beta^3)$ if $d$ is odd. This explains the "staircase" behavior of the correlation function  in Figure \ref{fig:correlation0}.

In the low temperature limit $\beta \to \infty$, the function $\alpha \to 1$  and the ratios $\lambda_2/\lambda_1$, $\lambda_4/\lambda_1$ behave as follows:
\begin{eqnarray}
\frac{\lambda_2}{\lambda_1} &=& 1-2\rme^{-4\beta} +\Or(\rme^{-6\beta}),\\
\frac{\lambda_4}{\lambda_1} &=& -\rme^{-2\beta} +\Or(\rme^{-4\beta}).\nonumber
\end{eqnarray}
In this limit, and in the large distance limit $d \to \infty$, the correlations decay exponentially as
\begin{eqnarray}
\left<s_i s_j\right>&=& \alpha \rme^{-d/\xi},
\end{eqnarray}
where $\xi=1/\log(\lambda_1/\lambda_2)$ is a correlation length. This behaviour is illustrated in Figure \ref{fig:correlation0}. In the limit $\beta \to \infty$, the correlation length $\xi$ diverges as $\xi \sim \frac{1}{2} \rme^{4\beta} $, identifying $T=0$ as the ``critical'' temperature of this system -- the system is in an ordered (ferromagnetic) state in the limit $T=0$.

It is interesting to compare this behaviour with the conventional one-dimensional Ising case, where $\xi_{\textrm{Ising}} \sim \frac{1}{2} \rme^{2\beta} $ \cite{Baxter1982}. This comparison shows, as could be expected, that adding a self-interaction, and at the same time increasing the size of the neighbourhood, helps to build correlations in this ferromagnetic system. In addition, the correlation function has at high temperature (and for small distances) an unusual ``staircase'' behaviour, which is probably due to the parallel dynamics. At lower temperatures, this effect disappears as competing spin domains become larger. It is also instructive to compare this result to the one-dimensional parallel Ising model presented in \ref{section:1dIsing}. A behaviour depending on the parity of the distance appears also there: the correlation function is zero for odd distances $d=j-i$ (at any temperature), and decreases exponentially with distance $d$ for even distances.

%

\section{Discussion\label{section:summary}}
In this paper, we have studied the 1d-MAJ3-CA (1-dimensional majority-3 cellular automaton) in the presence of thermal noise. This allows us to study properties of this automaton with methods of equilibrium statistical physics. In particular,  we first formulated the transfer matrix method for a more general 1-dimensional majority-(2k+1) automaton, where $k\geq1$. This method is exact and allows us to compute the moments of the equilibrium distribution (local magnetization, 2-point correlation, etc.) and various thermodynamic functions such as the free energy, the entropy, etc. for any system of finite size, but also in the thermodynamic limit.  

Applying the transfer matrix method to the 1d-MAJ3-CA with periodic boundary leads us to the conclusion that after an infinitely long time, the infinitely large 1d-MAJ3-CA forgets its initial configuration, at any positive (finite) temperature. This result contrasts with the result of~\cite{Mozeika2011} for non-uniform (random) topologies, where one bit of information can be remembered in the presence of noise. However, numerical experiments show that the 1d-MAJ3-CA has a quite long memory about its initial state.  The same results hold with open boundary -- the effect of the boundary is irrelevant in the thermodynamic limit. 

An interesting problem for future work is to study the 1d-MAJ(2k+1)-CA (1-dimensional majority-(2k+1) cellular automaton) model for any (finite) $k>1$. This is a difficult problem due to the exponential  growth of sizes of matrices used in the transfer matrix method.  In the extreme case where $k=N$, the majority automaton reduces to the infinite-range parallel-update Ising model~\cite{TheorOfIP}, which is fault-tolerant. We should also mention G\'{a}cs' construction of a one-dimensional CA~\cite{Gacs2001}, which is, to our knowledge, the only one-dimensional fault-tolerant CA known to date. Dynamical studies of the 1d-MAJ3-CA will answer the following questions: how quickly is memory of the initial state lost? And what are the effects of the self-interaction (in (\ref{def:process}) the state $s_i(t+1)$ is directly dependent on $s_i(t)$) on this property of the system? In systems with non-uniform topologies, the presence of self-interaction yields strong memory effects ~\cite{Mozeika2011}. A direct approach 
to this problem is to use the dynamical transfer matrix method~\cite{Coolen2012}.


\section*{Acknowledgements}
This work is supported by funding from the Center of Excellence
program of the Academy of Finland, with the COMP (251748) Centre for R\'emi Lemoy and the COIN (251170) Centre for Alexander Mozeika (AM). The work by Shinnosuke Seki is financially supported by HIIT Pump Priming Grant No.~902184/T30606 and by the Academy of Finland, Postdoctoral Research Grant No. 13266670/T30606. 
AM is thankful for interesting and helpful discussions with ACC Coolen and R K\"{u}hn.

\appendix

\section{Majority-$(2k+1)$ \label{section:Maj2k}}
For $k \ge 1$, the majority-$(2k+1)$ cellular automaton (maj-$(2k+1)$ CA) is a 1-dimensional array of $N$ cells $s_1, \ldots, s_{N}$ with binary states ($-1$ or $1$). Cells' states are updated synchronously by taking the majority vote among a cell, its $k$ left neighbours, and $k$ right neighbours. 

The partition function of maj-$(2k+1)$ CA has the following form: 
\begin{equation}\label{eq:partition_function}
	Z = \sum_{s_1, \ldots, s_N} \prod_{i = 1}^N 2 \cosh \left(\beta \mbox{$\sum_{j = -k}^{{+k}} s_{i+j}$} \right),  
\end{equation}
with $s_{-k+1} = s_{N-k+1}, \ldots, s_{k} = s_{N+k}$ (periodic boundary conditions). 

\begin{figure}
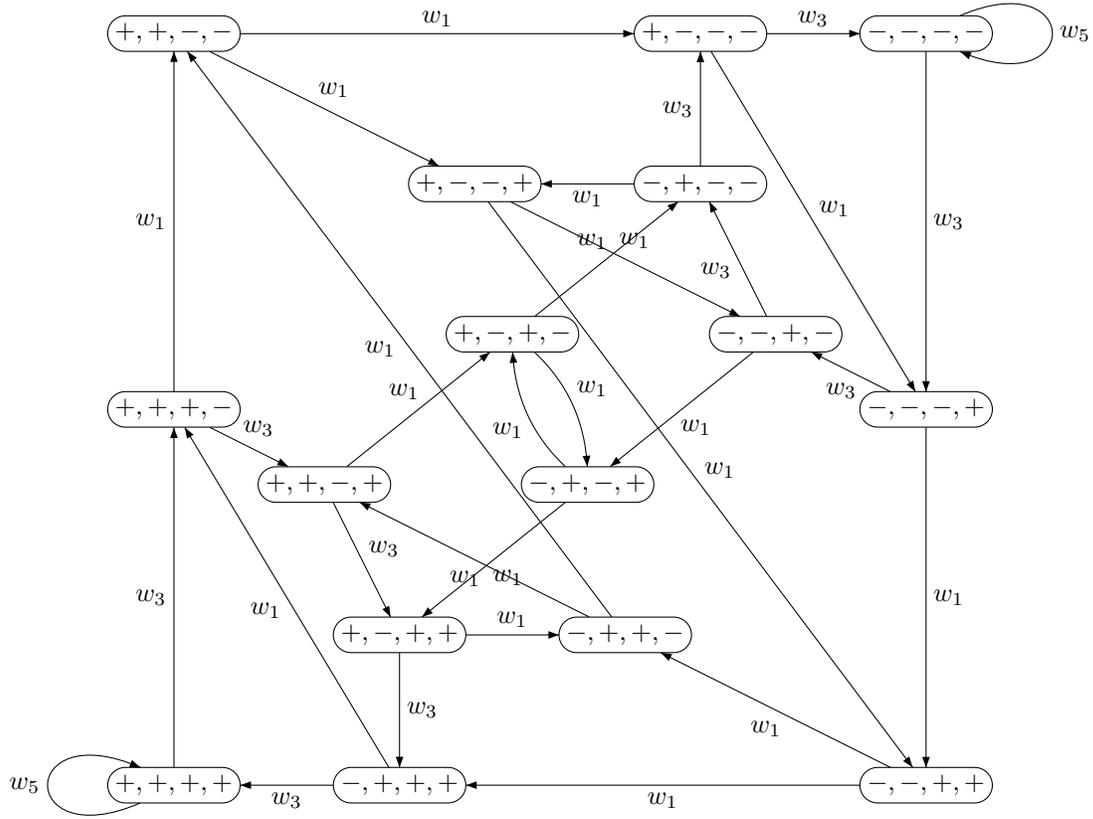

\begin{center}
\begin{gpicture}(100,120)

\node[Nadjust=wh](A1111)(0,0){$+, +, +, +$}

\node[Nadjust=wh](A1110)(0,50){$+, +, +, -$}
\node[Nadjust=wh](A1101)(20,40){$+, +, -, +$}
\node[Nadjust=wh](A1011)(30,20){$+, -, +, +$}
\node[Nadjust=wh](A0111)(30,0){$-, +, +, +$}

\node[Nadjust=wh](A1100)(0,100){$+, +, -, -$}
\node[Nadjust=wh](A1001)(40,80){$+, -, -, +$}
\node[Nadjust=wh](A0011)(100,0){$-, -, +, +$}
\node[Nadjust=wh](A0110)(60,20){$-, +, +, -$}

\node[Nadjust=wh](A0101)(55,40){$-, +, -, +$}
\node[Nadjust=wh](A1010)(45,60){$+, -, +, -$}

\node[Nadjust=wh](A0100)(70,80){$-, +, -, -$}
\node[Nadjust=wh](A0010)(80,60){$-, -, +, -$}
\node[Nadjust=wh](A1000)(70,100){$+, -, -, -$}
\node[Nadjust=wh](A0001)(100,50){$-, -, -, +$}

\node[Nadjust=wh](A0000)(100,100){$-, -, -, -$}

\drawloop[loopangle=180](A1111){$w_5$}\drawedge(A1111,A1110){$w_3$}
\drawedge(A1110,A1101){$w_3$}\drawedge(A1110,A1100){$w_1$}
\drawedge(A1101,A1011){$w_3$}\drawedge(A1101,A1010){$w_1$}
\drawedge(A1011,A0111){$w_3$}\drawedge(A1011,A0110){$w_1$}
\drawedge(A0111,A1111){$w_3$}\drawedge(A0111,A1110){$w_1$}

\drawedge(A1100,A1001){$w_1$}\drawedge(A1100,A1000){$w_1$}
\drawedge(A1001,A0010){$w_1$}\drawedge(A1001,A0011){$w_1$}
\drawedge(A0011,A0111){$w_1$}\drawedge(A0011,A0110){$w_1$}
\drawedge(A0110,A1101){$w_1$}\drawedge(A0110,A1100){$w_1$}
\drawedge[curvedepth=3](A1010,A0101){$w_1$}\drawedge(A1010,A0100){$w_1$}
\drawedge(A0101,A1011){$w_1$}\drawedge[curvedepth=3](A0101,A1010){$w_1$}

\drawedge(A0100,A1001){$w_1$}\drawedge(A0100,A1000){$w_3$}
\drawedge(A0010,A0101){$w_1$}\drawedge(A0010,A0100){$w_3$}
\drawedge(A1000,A0001){$w_1$}\drawedge(A1000,A0000){$w_3$}
\drawedge(A0001,A0011){$w_1$}\drawedge(A0001,A0010){$w_3$}

\drawloop[loopangle=0](A0000){$w_5$}\drawedge(A0000,A0001){$w_3$}

\end{gpicture}
\end{center}
\caption{Majority-5: transition diagram of the auxiliary 4-cell binary-state automaton, with $w_5 = 2\cosh(5\beta)$, $w_3 = 2\cosh(3\beta)$, and $w_1 = 2\cosh(\beta)$.}
\label{fig:maj5_graph}
\end{figure}
The transfer-matrix method used in this paper can be applied to the study of the maj-$(2k+1)$ CA. The auxiliary automaton, presented for $k=2$ (majority-5) in Figure \ref{fig:maj5_graph}, is then a $2k$-cell automaton.
In other words, the fundamental object is then the state of $2k$ consecutive cells of the maj-$(2k+1)$ CA. This state can have $2^{2k}$ different values, whose list we denote by $(\pm 1,\ldots,\pm 1)_{2k}$, and which we use as indices of the rows and columns of the transfer matrix $\bT$. This matrix has then $2^{2k}\times 2^{2k}$ elements $\bT[\tilde{s}_1,...,\tilde{s}_{2k}\vert \tilde{s}_{2k+1},...,\tilde{s}_{4k}]$, defined in the following way:
\begin{eqnarray}
&&\bT[\ldots\vert\ldots]=2\cosh \beta\left(\tilde{s}_1+\sum_{j=2k+1}^{4k} \tilde{s}_j \right)\prod_{\ell=1}^{2k-1}\delta_{\tilde{s}_{\ell+1};\tilde{s}_{2k+\ell}}
\end{eqnarray}
So that only $2^{2k+1}$ elements are non-zero. These non-zero elements are presented for $k=2$ by the diagram of Figure~\ref{fig:maj5_graph}. Then
\begin{eqnarray}
\Tr(\bT^N) &=\sum_{a_1,\ldots,a_N=(\pm 1,\ldots,\pm 1)_{2k}} \prod_{i=1}^N \bT_{a_i, a_{i+1}}\\
&= \sum_{s_1,...,s_N}  \prod_{i=1}^N \bT[  s_{i-k},\ldots,s_{i+k-1}\vert s_{i-k+1},\ldots,s_{i+k}]\nonumber\\
&=Z\nonumber
\end{eqnarray}
where $a_{N+1}=a_1$ (periodic boundary), and the second equality is due to the properties of $\bT$.

Unfortunately, the transfer matrix has a size growing exponentially with $k$, which makes it difficult to diagonalize, even for $k=2$. However, numerical treatment seems to indicate that in this case it still has distinct real eigenvalues of multiplicity one.

\section{Matrix $\bM$ \label{mat}}
The matrix $\bM=\bP^{-1}\bS\bP$ defined in section \ref{section:equilibrium} has the form
\begin{equation}
\bM=
\left(
\begin{array}{cccc}
 0 & \bM_{1,2} & 0 & \bM_{1,4} \\
\bM_{2,1} &  0 & \bM_{2,3} & 0 \\
0 & \bM_{3,2} & 0 & \bM_{3,4}  \\
\bM_{4,1} &  0 & \bM_{4,3} & 0  \\
\end{array}
\right).
\end{equation}
Defining 
\begin{equation}
f(\beta)= \cosh (\beta) \sqrt{2\cosh (4 \beta)-8 \cosh (2 \beta)+10},
\end{equation}
we can write the non-zero elements of $\bM$ as
{
\begin{eqnarray}
\bM_{1,2} =&\frac{ \sinh (2\beta) (\cosh (2 \beta)+\sinh (2 \beta)) }{2f(\beta) (\cosh (\beta)+\sinh (\beta))}\times\nonumber \\
&\times\big(-4 \cosh (3 \beta)+2 \sinh (\beta)+2 \sinh (3 \beta)-2 f(\beta)\big),\nonumber \\
 \bM_{1,4}=& -\frac{\sinh (2\beta) (\cosh (\beta)+\sinh (\beta)) (\cosh (2 \beta)-\sinh (2 \beta)) }{f(\beta)}\times\nonumber \\
&\times\big(2 \cosh (3 \beta)+\sinh (\beta)+\sinh (3 \beta)+f(\beta)\big),\nonumber \\
\bM_{2,1}=& -\frac{2 \cosh (\beta) (\cosh (\beta)+\sinh (\beta))^2 }{\cosh (\beta)+\cosh (3 \beta)+f(\beta)}\times\nonumber \\
&\times \big(2 \cosh (\beta)-3 \sinh (\beta)+\sinh (3 \beta)+f(\beta)\big), \nonumber \\
\bM_{2,3}=& -\frac{2 \cosh (\beta) (\cosh (\beta)+\sinh (\beta))^2 }{\cosh (\beta)+\cosh (3 \beta)-f(\beta)}\times\nonumber \\
&\times \big(2 \cosh (\beta)-3 \sinh (\beta)+\sinh (3 \beta)-f(\beta)\big), \nonumber \\
\bM_{3,2}=& -\frac{\sinh (2\beta) (\cosh (2 \beta)+\sinh (2 \beta)) }{f(\beta) (\cosh (\beta)+\sinh (\beta))}\times\nonumber \\
&\times\big(-2 \cosh (3 \beta)+\sinh (\beta)+\sinh (3 \beta)+f(\beta)\big),\nonumber \\
\bM_{3,4}=& \frac{(\cosh (\beta)-\sinh (\beta)) \sinh (2 \beta) }{f(\beta)}\times\nonumber \\
&\times \big(2 \cosh (3 \beta)+\sinh (\beta)+\sinh (3 \beta)-f(\beta)\big), \nonumber \\
\bM_{4,1}=& \frac{\cosh (\beta) (\cosh (\beta)-\sinh (\beta)) }{\big(\cosh (\beta)+\cosh (3 \beta)+f(\beta)\big) (\cosh (\beta)+\sinh (\beta))}\times\nonumber \\
&\times \big(4 \cosh (\beta)+6 \sinh (\beta)-2 \sinh (3 \beta)+2 f(\beta)\big), \nonumber \\
\bM_{4,3}=& \frac{\cosh (\beta) (\sinh (2 \beta)-\cosh (2 \beta)) }{\cosh (\beta)+\cosh (3 \beta)-f(\beta)}\times\nonumber \\
&\times \big(-4 \cosh (\beta)-6 \sinh (\beta)+2 \sinh (3 \beta)+2 f(\beta)\big). \nonumber \\
\end{eqnarray}
}

\section{One-dimensional parallel Ising model\label{section:1dIsing}}

The results presented in this work can be also compared with the correlations between cells in the (one-dimensional) parallel Ising model under thermal noise, in which the cell does not refer to itself in the update of its state. 
Under the noise distribution (\ref{def:noise}), the parallel Ising model has a unique equilibrium probability distribution
\begin{eqnarray}
  \Prob_{\infty}(s)= \frac{1}{Z} \prod_{i=1}^N 2\cosh(\beta(s_{i-1}+s_{i+1})), \label{def:1dIsing}
\end{eqnarray}
where $Z = \sum_s \prod_{i=1}^N  2\cosh(\beta(s_{i-1}+s_{i+1}))$. 
In order to compute the correlation function $\langle s_i s_j \rangle$, let us rewrite (\ref{def:1dIsing}) as 
\begin{eqnarray}
	\Prob_{ \infty}(s) &=& \frac{1}{Z}   \prod_{i=1}^N     \exp(\ln(\cosh(\beta(s_{i-1} + s_{i+1})))) \label{eq:1dIsing} \\
	&=& \frac{1}{Z} \prod_{i=1}^N \exp \left(\frac{1}{2}(1 + s_{i-1} s_{i+1})\ln\cosh(2\beta)\right) \nonumber \\
&=& \rme^{\tilde{\beta}N} \frac{1}{Z_{\rm even}}\exp\left(\tilde{\beta}\sum_{i} s_{2i} s_{2i+2} \right) \frac{1}{Z_{\rm odd}} \exp \left(\tilde{\beta}\sum_{i} s_{2i-1} s_{2i+1} \right), \nonumber
\end{eqnarray}
where $\tilde{\beta} = \frac{1}{2}\ln \cosh(2\beta)$. The factorization in (\ref{eq:1dIsing}) suggests that the parallel Ising model is actually equivalent to two independent sequential Ising models operating at temperature $1/\tilde{\beta}$. 
For a sequential Ising model at temperature $1/\tilde \beta$, the correlation between cells at distance $d=|i-j|$ is given by $\tanh(\tilde \beta)^d$~\cite{Baxter1982}. 
Thus, the correlation $\langle s_i s_j \rangle$ between cells of the parallel Ising model is equal to $
%
	\langle s_i s_j \rangle = 
	\left(\frac{\cosh(2\beta)-1}{\cosh(2\beta)+1}\right)^{|i-j|} 
%
$
when  $d=|i-j|$ is even, and it is zero if $d$ is odd. 

\section*{References}



\end{document}